\documentclass[fleqn,usenatbib]{mnras}

\usepackage{newtxtext,newtxmath}
\usepackage[T1]{fontenc}
\usepackage{ae,aecompl}

\usepackage{graphicx}	% Including figure files
\usepackage{amsmath}	% Advanced maths commands
\usepackage{amssymb}	% Extra maths symbols

\defcitealias{simard2011}{S11}
\defcitealias{leslie2016}{L16}
\title[Dissecting the Main Sequence]{Dissecting the Main Sequence: AGN Activity and Bulge Growth in the Local Universe}

\author[C. McPartland et al.]{
C. McPartland$^1$\thanks{E-mail: conormcp@ifa.hawaii.edu (CM)}
D. B. Sanders$^1$, L. J. Kewley$^2$, S. K. Leslie$^3$
\\
$^1$Institute for Astronomy, 2680 Woodlawn Dr., Honolulu, HI 96822, USA \\
$^2$Research School of Astronomy \& Astrophysics, Australian National University, Cotter Road, Weston, ACT 2611, Australia\\
$^3$Max-Planck Institut fur Astronomie, Koningstuhl 17, D-69117, Heidelberg, Germany
}

\date{Accepted 2018 October 19. Received 2018 October 3; in original form 2018 June 30}

\pubyear{2018}

\begin{document}
\label{firstpage}
\pagerange{\pageref{firstpage}--\pageref{lastpage}}
\maketitle

\begin{abstract}
Local galaxies from the Sloan Digital Sky Survey are used to provide additional support for an evolutionary pathway in which AGN activity is associated with star-formation quenching. Composite, Seyfert 2 and LINER galaxies account for $\sim$60\% of all star-formation in massive galaxies ($M_\star > 10^{10.5} M_\odot$).  Inclusion of these galaxies results in a ``turnover" in the $SFR - M_\star$ relation for massive galaxies. Our analysis shows that bulge growth has already occurred in the most massive galaxies ($M_\star > 10^{10.5}$ $M_\odot$), and bulges continue to grow as galaxies quench  and redden, $(g-r)$ = 0.5{\thinspace}$\rightarrow${\thinspace}0.75. Significant bulge growth is also occurring in low mass starburst galaxies ($M_\star < 10^{10.5} M_\odot$) at 0.5{\thinspace}dex above the "main sequence" (MS), where we find an increase in $B/T$ from 0.1{\thinspace}$\rightarrow${\thinspace}0.3 and bluer colours, $(g-r) < 0.25$ compared to low-mass galaxies on the MS. 
\end{abstract}


%%%%%%%%%%%%%%%%%%%%%%%%%%%%%%%%%%%%%%%%%%%%%%%%%%

%%%%%%%%%%%%%%%%% BODY OF PAPER %%%%%%%%%%%%%%%%%%

\section{Introduction}

Over the past decade, it has become clear that the star-formation rates ($SFR$) and stellar masses ($M_\star$) of star-forming galaxies are highly correlated. Early studies revealed the existence of a roughly linear $SFR$--$M_\star$ relation for star-forming galaxies to at least $z\sim2$; and gave the relation the name the star-forming galaxy ``main sequence'' \citep[hereafter the MS;][]{noeske2007,daddi2007,elbaz2007}. The salient features of the MS from these early studies are that normalization of the MS increases with redshift, while the intrinsic scatter in the relation appeared to be small (0.2 - 0.3 dex) at all redshifts studied \citep{speagle2014}. It was argued that the existence of the MS implied that star-formation was regulated by a limited set of internal processes, (e.g. gas supply), followed by an additional unknown process that ``quenches'' the galaxy (e.g. gas exhaustion); transforming the galaxy in colour from blue to red. 

\citet{oemler2017} demonstrate that most star-formation history models will produce a tight main sequence, and thus yield little astrophysical insight. However, the properties of the quiescent population require additional astrophysical processes to explain. Early studies, which generally excluded AGN from their analysis, argued that bulge size and local environmental density were the best predictors of quenching \citep[e.g.][]{cheung2012,fang2013,omand2014}. With the availability of data from the Spitzer and Herschel space telescopes, studies using large samples of robust mid- and far-IR based SFR measurements uncovered a flattening, or ``turnover'', in the slope of the MS above $M_\star =\sim3\times10^{10} M_\odot$ in local galaxies \citep{salim2007} as well as at higher redshifts \citep{whitaker2014,lee2015,tomczak2016}. 

The existence of a turnover in the MS implies that massive galaxies have lower average SFRs than would be expected if there were a linear MS. \citet{abramson2014} argue that the ``turnover" in the MS may be a result of the increased bulge mass-fractions in massive galaxies, which causes the total $M_\star$ to be a poor proxy for the star-forming gas, which primarily resides in the disk. More recently, \citet[hereafter L16]{leslie2016} have reanalyzed the SFR data for SDSS galaxies by galaxy spectral type, and have shown that Composites, Seyfert 2, and LINER galaxies form a "quenching pathway" for massive galaxies, which is nearly perpendicular to the star-forming MS, suggesting that active galactic nuclei (AGN) may play a key role in quenching star formation and removing massive galaxies from the MS. 

% Roadmap for Paper
In this \emph{letter}, we continue to use SDSS data to further explore the nature of the ``turnover" in the MS due both to AGN and bulge growth in massive galaxies.  Specifically, we use the subset of the \citetalias{leslie2016} sample with bulge+disk photometry from \citet[hereafter S11]{simard2011} to address two main questions: 1) Which galaxies produce the turn over in the MS at high $M_\star$? Is the ``turnover'' a result of sample biases, or an indication of a different mechanism regulating star-formation in massive galaxies?, and 2) What makes a disk galaxy grow a bulge, and how is the growth of that bulge related to AGN activity and quenching? The data sample and spectral classification methods are reviewed in \S{2}. The results are presented in \S{3}, and discussed in \S{4}. A summary of our analysis and conclusions are given in \S{5}. 

\section{Data Sample}\label{sec:data}
Our galaxy sample is drawn from the SDSS data release 7 \citep{abazajian2009}. Redshifts from the \citetalias{simard2011} bulge+disk decomposition catalogue are used to select galaxies between redshifts $z$ = 0.04 and 0.1 (as in \citetalias{leslie2016}). For each of the selected galaxies, we extract the $r$-band bulge-to-total fraction, $B/T$ and total $gr$ absolute magnitudes, $M_{g,g}$, $M_{r,g}$. We use the ``n4''-model from \citetalias{simard2011} which assumes a de Vaucouleurs/exponential profile for the bulge/disk, and is more robust than the ``fn''-model which uses a S\`ersic profile for the bulge. We considered using stellar mass $B/T$ ratios from the catalog of \citet{mendel2014}, but found that the average errors on these measurements are too large for the analysis presented in this letter.

Star-formation rates, stellar masses, and emission line fluxes are extracted from the MPA/JHU SDSS DR7 catalogues\footnote{\url{http://www.mpa-garching.mpg.de/SDSS/DR7/}}. Star-formation rates in the MPA/JHU catalogue are measured using the method \citep{brinchmann2004} which relies on dust-corrected H$\alpha$ luminosities for star-forming galaxies, and an empirical relation between D4000 and specific $SFR$ ($SFR/M_\star$) for the remaining sources. Stellar masses estimates are based on fits to the photometry following \citet{kauffmann2003}. Here, we use the catalogue version $5.2b$, but note little difference from the original measurements. Aperture corrections in the MPA/JHU catalogues are applied to $SFR$s and stellar mass estimates following \citet{salim2007}. We adopt the median estimates for $SFR$ and $M_\star$. The entire analysis was repeated using $SFR$ estimates from \citet{ellison2016}, as well as $SFR$ and $M_\star$ estimates from the Galex-SDSS-WISE Legacy Catalogue \citep[GSWLC:][]{salim2016}. Here, we adopt the MPA/JHU measurements since they provide the largest galaxy sample, and a direct comparison with studies already in the literature.

There are 290,976 sources at $z=0.04-0.1$ with measurements of [$M_\star$, $SFR$] from the MPA/JHU catalogue, and [$B/T$, $M_{g,g}$, $M_{r,g}$] from the \citetalias{simard2011} catalogue. Since spectral classification is essential to our analysis, following \citetalias{leslie2016}, passive galaxies (i.e. sources with  H$\alpha$ $\mathrm{S/N}$ < 3) are removed from the BPT analysis. This results in a final sample of 203,162 galaxies in the desired redshift interval with satisfactory measurements of $SFR$, $M_\star$, $B/T$, $(g-r)$ colour, and emission line ratios. 

\subsection{Spectral Classification}\label{sec:classifcation}
\begin{figure}
 \centerline{
  % CLASS FRACTION HISTOGRAM
  \includegraphics[width=0.5\textwidth]{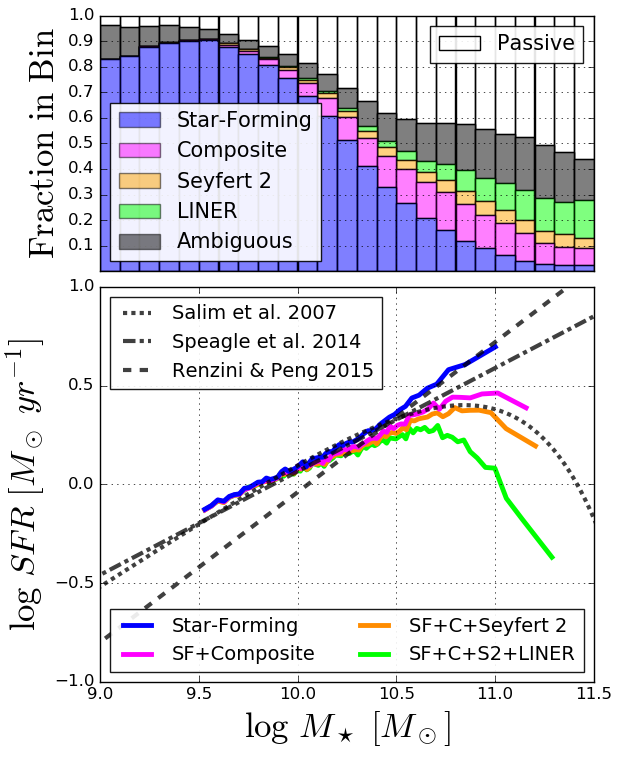}
 }
 \caption{
  \emph{top:} The fraction of galaxies classified as Star-Forming, Composite, Seyfert 2, LINER, ambiguous, and passive in bins of stellar mass.
   \emph{bottom:} The median star-formation rate as a function of stellar mass for galaxies in four subsets of spectral classes (coloured lines). The dashed and dotted lines show measurements from the studies indicated at the top left.
   \label{fig:histMS}
 }
\end{figure}

Following \citetalias{leslie2016}, we adopt the \citet{kewley2006} classification scheme, which builds on diagnostic diagrams proposed by \citet{BPT,veilleux1987} to classify to the dominant source of ionizing flux in an emission line galaxy. We use the emission line ratios [\ion{N}{ii}]/H$\alpha$, [\ion{S}{ii}]/H$\alpha$, [\ion{O}{i}]/H$\alpha$, and [\ion{O}{iii}]/H$\beta$ to divide our sample into four spectral classes: Star-Forming, Composite, Seyfert 2, and LINER galaxies. We exclude ambiguous galaxies -- i.e. those with conflicting classifications -- from the BPT analysis (e.g. Composite in [\ion{O}{iii}]/H$\beta$--[\ion{N}{ii}]/H$\alpha$ but Seyfert 2 in [\ion{O}{iii}]/H$\beta$--[\ion{S}{ii}]/H$\alpha$). The ambiguous galaxies are included in Fig.~\ref{fig:histMS} to simply provide an accurate measure of the passive galaxy fraction.

\begin{figure*}
 \centerline{
  % NUMBER DENSITY
  \includegraphics[clip, trim=0mm 0mm 5mm 15mm, width=0.95\textwidth]{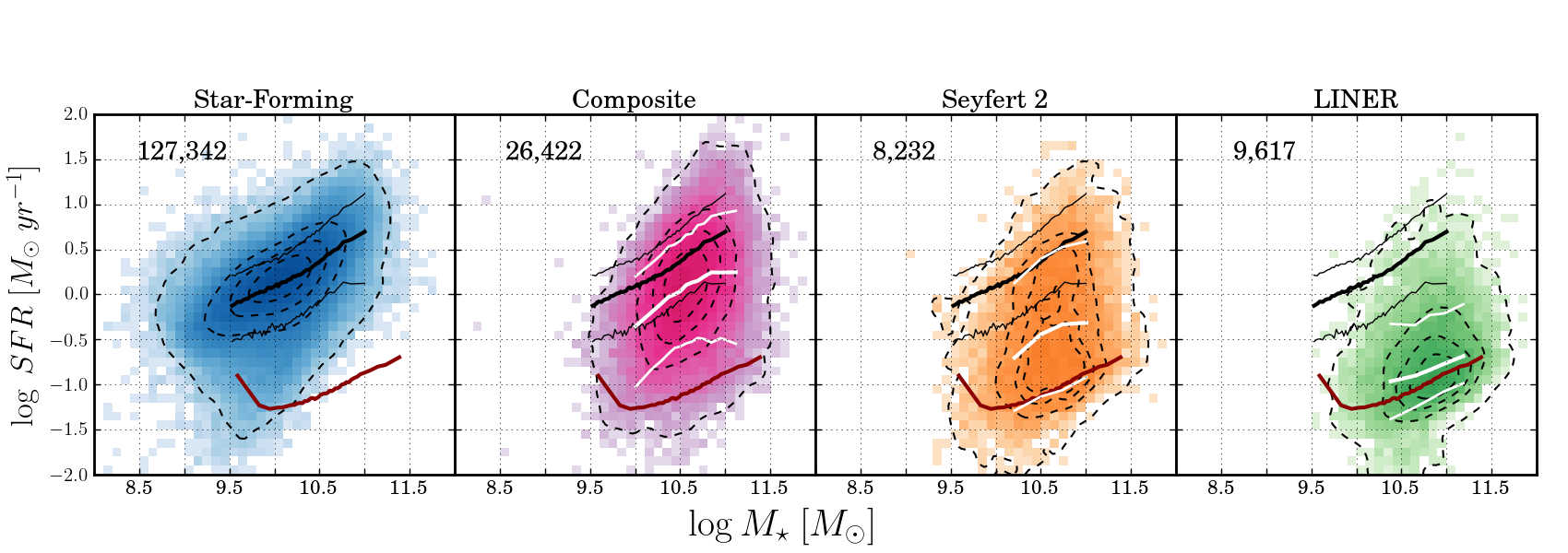}
 }
 \caption{The logarithmic number density of galaxies by spectral class in 40$\times$40 bins of $\Delta SFR \times \Delta M_\star$. The top of each panel shows the name and sample size of the respective class. The dashed contours encompass 25, 50, 90, 99 percent of the sources in each panel. The solid black lines indicate the median and 10/90\% quantiles in $SFR$ as a function of $M_\star$ for Star-Forming galaxies. The white lines show the same for the other three classes. The red line indicates the median $SFR$ for passive galaxies.\label{fig:sfr_mstar}}
\end{figure*}

\section{The \texorpdfstring{$SFR$--$M_\star$}{SFR--M*} diagram by spectral class}
\label{sec:ms}
In this section, we present the distribution of each spectral class in the $SFR$--$M_\star$ diagram. For each spectral class, we divide galaxies into 0.1$\times$0.1 dex bins of $\Delta\log SFR\times\Delta\log M_\star$. In Fig.~\ref{fig:sfr_mstar}, the colour scale indicates the logarithm of the number of galaxies in each bin. The panels from left to right show Star-Forming, Composite, Seyfert 2, and LINER galaxies. The numbers in each panel indicate the total number of galaxies in our sample with the corresponding spectral classification. The dashed contours show the 25,50,90,99\% quantiles of the Gaussian smoothed joint distribution in $SFR$--$M_\star$ for the respective spectral class. 

The $SFR$--$M_\star$ relation for Star-Forming galaxies provides a useful benchmark to understand how other spectral classes differ. The solid black lines in each panel of Fig.~\ref{fig:sfr_mstar} indicate the median and 10,90\% quantiles in $SFR$ as a function of $M_\star$ for Star-Forming galaxies (see \S~\ref{sec:msturn}, same as the blue line in Fig.~\ref{fig:histMS}); the white lines show the same for Composite, Seyfert 2, and LINER galaxies. The red line indicates the median $SFR$ for passive galaxies in our sample (see \S~\ref{sec:data}). The median $SFR$ for Composite galaxies with stellar masses $\gtrsim10^{10.5}\ M_\odot$ is a factor of $\sim$2 lower than Star-Forming galaxies in the same mass interval. However, the Composite galaxy distribution has broad tails that extend up to the highest $SFR$s in our sample, and down to $SFR$s consistent with the quenched population. Seyfert 2 galaxies above $10^{10.5}\ M_\odot$ have a median $SFR$ an additional factor $\sim$3 lower than Composite galaxies, but have a similar median stellar mass of $M_\star \sim 6\times10^{10}\ M_\odot$. LINER galaxies have a slightly larger median stellar mass around $M_\star \sim 7.5\times10^{10}\ M_\odot$, with an additional factor of $\sim$3 decrease in SFR compared to Seyfert 2 galaxies. 

\subsection{The turnover in \texorpdfstring{$SFR$--$M_\star$}{SFR--M*} at high \texorpdfstring{$M_\star$}{M*}}
\label{sec:msturn}
We split galaxies into bins of $\Delta \log M_\star$, and show the fraction of galaxies from each spectral type in the top panel of Fig.~\ref{fig:histMS}. For completeness, we include passive and ambiguous galaxies (see \S~\ref{sec:data}~\&~\ref{sec:classifcation}). Star-Forming galaxies represent the majority of the galaxy population at lower stellar masses, but galaxies classified as Composite, Seyfert 2, and LINER outnumber them above a stellar mass of $M_\star \sim 10^{10.6}\ M_\odot$. We next investigate how including these ``non-Star-Forming'' galaxies influences measurements of the MS.

To determine the shape of the MS, we compute the median $SFR$ as a function of $M_\star$ using natural width bins of 2000 galaxies each. We begin with only Star-Forming galaxies and then repeat the calculation after adding Composite, Seyfert 2, and LINER galaxies. We show the results of this exercise in the bottom panel of Fig.~\ref{fig:histMS}. The MS considering only Star-Forming galaxies is roughly linear with slope and normalization consistent with \citet{renzini_peng2015}, as well as \citet[after converting back to a Kroupa IMF]{elbaz2007}. Adding Composite and Seyfert 2 galaxies flattens the MS at high stellar masses producing a relation consistent with the MS for all galaxies found by \citet{salim2007}, and similar to other results at higher redshifts \citep[e.g.][]{whitaker2014,lee2015,tomczak2016}.

The growth of a central bulge could also produce the high-mass turnover in the MS. \citet{abramson2014} show the local MS straightens if one removes the mass of the bulge from the total mass. At higher redshifts, bulge dominated galaxies have also been shown to flatten the high mass end of the $SFR$--$M_\star$ relation \citet{whitaker2015,erfanianfar2016}. However, since supermassive blackholes and bulges maintain an approximately constant mass ratio, many authors argue that the anti-correlation between bulge-fraction and $SFR$ supports a paradigm in which AGN feedback quenches star-formation \citep{bell2012,bluck2014,bluck2016,lang2014,teimoorinia2016,brennan2017}. In the next section, we investigate how both bulge fraction \emph{and} AGN activity relate to the $SFR$s of local galaxies.

\begin{figure*}
 % BULGE/TOTAL
 \centerline{
  \includegraphics[clip, trim=0mm 13mm 0mm 0mm, width=0.95\textwidth]{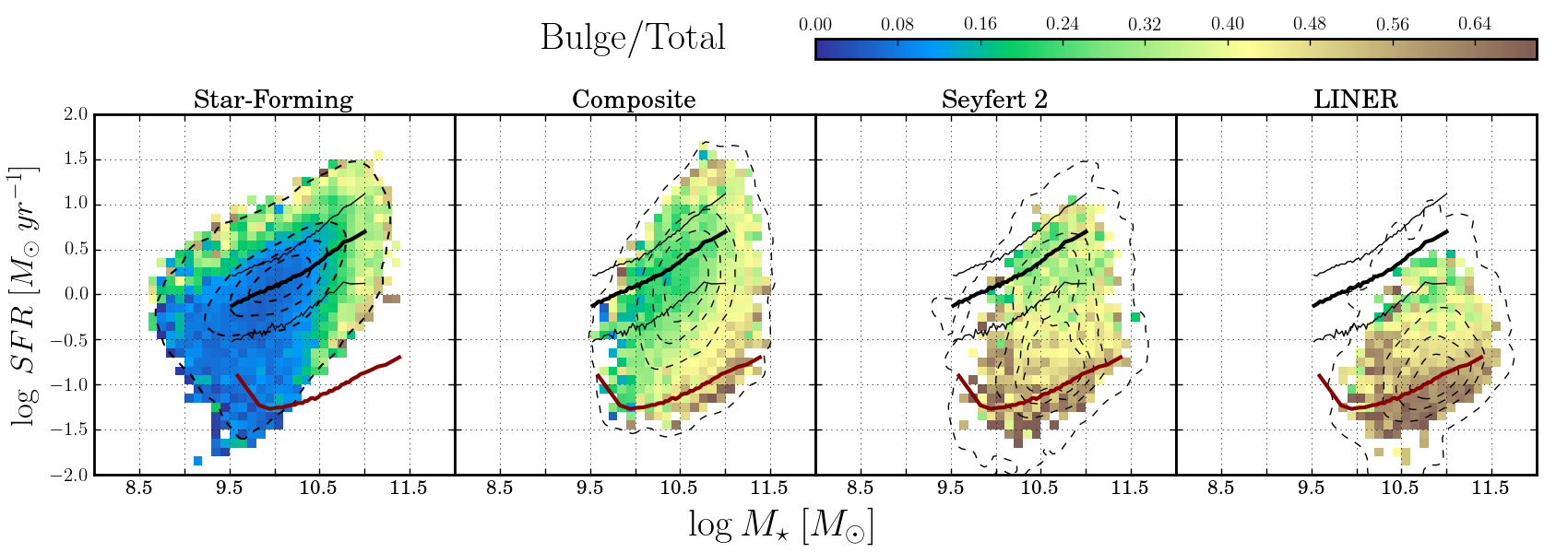}
  }
 % BULGE COLOR
 \centerline{
  \includegraphics[clip, trim=0mm 0mm 0mm 0mm,  width=0.95\textwidth]{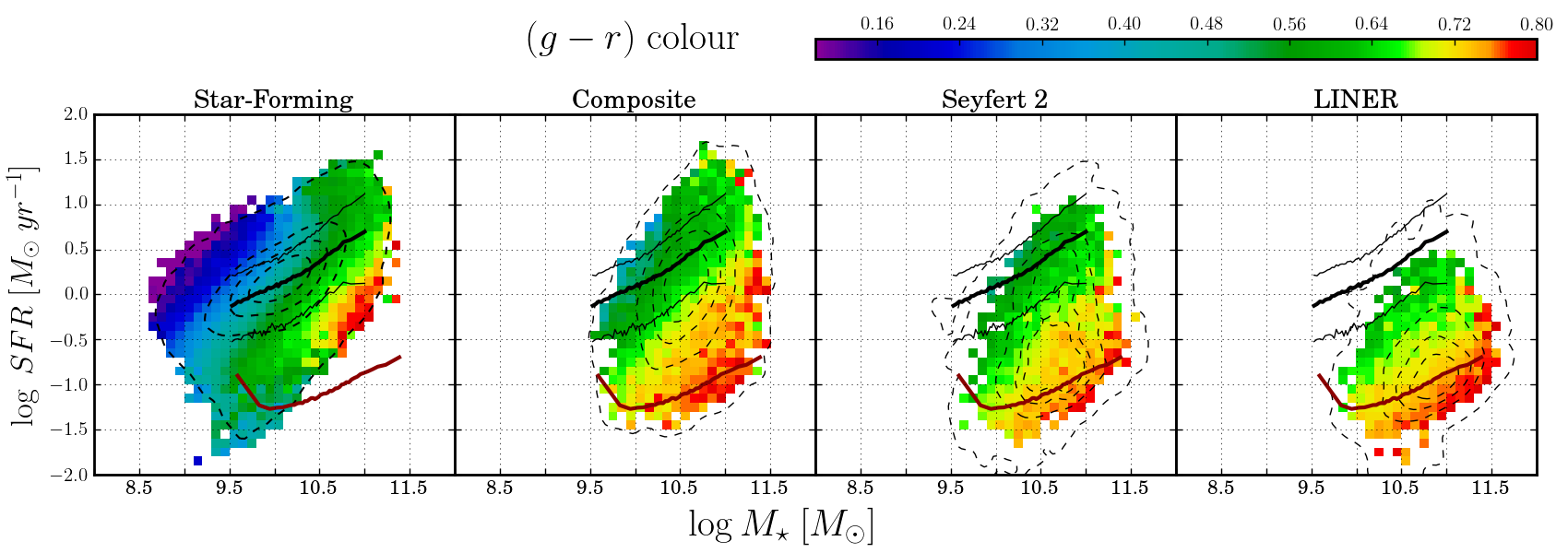}
 }
 \caption{The median bulge-to-total fraction, $B/T$ (top), and $(g-r)$ colour (bottom) using the same bins as in Fig.~\ref{fig:sfr_mstar}. Here, we reproduce the contours, and median/quantile lines from Fig.~\ref{fig:sfr_mstar} (although we omit the white lines for clarity). Note that the $B/T$ measurements for individual galaxies do indeed range from 0 to 1. However, we limit the $B/T$ colour range to 0--2/3 to highlight the dynamic range of the medians. \label{fig:bulges}}
\end{figure*}

% BULGE/TOTAL
\section{Bulge/Total and Galaxy Colour in the \texorpdfstring{$SFR$-$M_\star$}{SFR--M*} Diagram}
\label{sec:bt}
We present the bulge/total fractions and galaxy colours of each spectral class in the $SFR$--$M_\star$ diagram. We calculate the median $B/T$, and galaxy $(g-r)$ colours in each of the bins used in Fig.~\ref{fig:sfr_mstar}. The top and bottom panels of Fig.~\ref{fig:bulges} show the result of this calculations for $B/T$ and galaxy $(g-r)$ colour. The colour scale in the top panel indicates the median $B/T$ with dark blue corresponding to a pure-disk galaxy with $B/T=0$, and brown corresponding to a bulge-dominated galaxy with $B/T=2/3$. The colour scale in the bottom panel of Fig.~\ref{fig:bulges} indicates the median rest-frame $(g-r)$ colour for the galaxies in each bin. Following definitions from \citet{mendel2014}, who used a similar sample of SDSS galaxies, \emph{purple} corresponds with $(g-r)$ = 0.1, the colour of the bluest galaxies in the Blue Cloud; \emph{red} corresponds with $(g-r)$ = 0.8, the colour of the Red Sequence at $M_\star \sim 10^{11}\ M_\odot$; and \emph{green} corresponds with $(g-r)$ = 0.6, the colour of the Green Valley at $M_\star \sim 10^{10}\ M_\odot$. The dashed contours and solid lines (black and red) are the same as in Fig.~\ref{fig:sfr_mstar}.

There is a sharp $\sim$30--40\% increase in the median $B/T$ for Star-Forming galaxies with star-formation rates $\gtrsim$3 times higher than the median for their stellar mass. There is a similarly sharp increase in $B/T$ at all $SFR$s above a stellar mass of $M_\star \sim 10^{10.5}\ M_\odot$. \citet{wuyts2011b} find similar trends in the S\'esic indices of local SDSS galaxies. The medians in $B/T$ for the remaining spectral classes are $\gtrsim$30\% throughout most the $SFR$--$M_\star$ diagram.  Bulge fractions for Composite galaxies increase above the MS, but peak at the lowest SFRs and highest masses. The median $B/T$ of Seyfert 2 and LINER galaxies increases with decreasing $SFR$, as well with decreasing stellar mass. This is the opposite trend than what we find for Star-Forming galaxies, and could possibly be due to emission from an AGN that has not been accounted for in the \citetalias{simard2011} bulge+disk decompositions. Deeper imaging data with a smaller PSF FWHM than SDSS will help to disentangle emission in the bulge from that of a compact nucleus.

The $(g-r)$ colours of Star-Forming galaxies with stellar mass $M_\star\lesssim 10^{10.5}\ M_\odot$ in general correlate inversely with specific star-formation rate (i.e. $SFR/M_\star$). The exception being low-mass low-$SFR$ galaxies with $M_\star \sim 10^{9.5} M\odot$ and $SFR \sim 0.03-0.1\ M_\odot yr^{-1}$ where the median $(g-r)$ colours are similar to MS galaxies of the same mass. However, this may be an artefact of incompleteness in the SDSS spectroscopic sample at these low masses and star-formation rates. More massive Star-Forming galaxies above the MS have green median colours of $(g-r) \sim 0.6$; consistent with ``Green Valley'' galaxies. The median colours of massive Star-Forming galaxies with the lowest SFR are consistent with the Red Sequence despite featuring Milky Way level star formation rates of $SFR \sim 1\ M_\odot$ yr$^{-1}$. 

Composite and Seyfert 2 galaxies with $SFR$ greater than the median for their mass and spectral type also feature median colours that would place them in the Green Valley. LINER galaxies in the top 10th percentile in SFR have similar colours as well. Red Composite galaxies have a broader distribution in $SFR$ than either Seyfert 2 or LINER galaxies; suggesting Composite galaxies may have more dust, and thus more reddening, than other galaxies of similar $SFR$ and $M_\star$. Seyfert 2 and LINER galaxies get redder as specific $SFR$ decreases, with a weak trend toward bluer colours with decreasing stellar mass.

%%%%%%%%%%%%%%%%%%%
%%  CONCLUSIONS  %%
%%%%%%%%%%%%%%%%%%%
\section{Conclusions}
% Summary
We use SDSS-DR7 galaxies with bulge+disk decomposition from \citetalias{simard2011} to investigate how the bulge/total ($B/T$) ratio of galaxies varies as a function of galaxy spectral type and location in the $SFR-M_\star$ diagram. We also show how the inclusion of star-formation from galaxies of different spectral types affects the mean $SFR-M_\star$ relation at high $M_\star$.   Our main findings are given below.  

% Conclusions
\begin{enumerate}
\item For massive galaxies ($M_\star > 10^{10.5} M_\odot$), the fraction classified as either Composite, Seyfert 2 or LINER increases systematically with increasing $M_\star$, reaching $>$50\% at $M_\star > 10^{10.5} M_\odot$; therefore, samples selected to contain only galaxies classified as ``pure star-forming'' miss a significant fraction of the massive galaxy population.  Nearly 60\% of the total star formation in galaxies with $M_\star \gtrsim 10^{10.5}  M_\odot$ occurs in galaxies classified as non-star-forming (i.e. Composite, Seyfert 2, or LINER).  
 \item Galaxies classified as Composite cover a broad range in SFR, extending up to 1{\thinspace}dex above the MS  (consistent with powerful starbursts), and up to 2{\thinspace}dex below the MS (consistent with passive galaxies). Including Composite galaxies when computing the median $SFR(M_\star)$ flattens the $SFR-M_\star$ relation above $M_\star=10^{10.5} M_\odot$ and creates a negative slope (i.e. "turnover") above $M_\star=10^{11} M_\odot$.  Further inclusion of Seyfert 2 and LINER galaxies pushes the the turnover point to lower $M_\star$.
\item  For low-mass star-forming galaxies ($M_\star < 10^{10.5} M_\odot$), the median $B/T$ increases sharply (0.1{\thinspace}$\rightarrow${\thinspace}0.3) at $SFR \sim$0.5{\thinspace}dex above the MS.  These galaxies have bluer colours, $(g-r) < 0.25$, than their MS counterparts, $(g-r) \sim 0.3 - 0.5$. 
\item All massive galaxies ($M_\star > 10^{10.5} M_\odot$) have median $B/T > 0.3$ and green to red colours, $(g-r) > 0.5$. Composite galaxies show a clear trend of increased $B/T$ ($> 0.4$) above $M_\star \sim 10^{11}$. The largest bulges, $B/T > 0.5$, and the reddest colors, $(g-r) > 0.75$, are both found in galaxies with $SFR \gtrsim 1.5${\thinspace}dex below the MS.  
\end{enumerate}

Our results further demonstrate that AGN activity is associated with star-formation quenching in massive galaxies, and strongly suggest that bulge growth is an integral part of this quenching pathway. Future work with deeper and higher resolution imaging data will be necessary to better address the possible connections between AGN and bulge growth in quenching star-formation. The Pan-STARRS 1 (PS1) 3$\pi$ survey \citep{chambers2016} will allow better decomposition of galaxy radial profiles \citep{lokken2018}. Large spatially resolved integral field spectroscopy data sets, such as those provided by MaNGA \citep{bundy2015} and SAMI \citep{croom2012} will also help to shed light on this issue by allowing studies of the spatial distribution of star-formation in relation to BPT classifications.

\section*{Acknowledgements}
The authors would like to thank the anonymous referee for the very helpful comments that greatly improved the quality of the manuscript.
Funding for the SDSS and SDSS-II has been provided by the Alfred P. Sloan Foundation, the Participating Institutions, the National Science Foundation, the U.S. Department of Energy, the National Aeronautics and Space Administration, the Japanese Monbukagakusho, the Max Planck Society, and the Higher Education Funding Council for England. The SDSS Web Site is http://www.sdss.org/.    D.S. would like to thank he Distinguished Visitor Program at the Research School for Astronomy and Astrophysics, Australian National University for their generous support while in residence at the Mount Stromlo Observatory, Weston Creek, ACT.  C.M. and D.S. also acknowledge support from NSF grant number 1716994. The work of D.S. was performed in part at the Aspen Center for Physics, which is supported by National Science Foundation grant PHY-1607611.

%%%%%%%%%%%%%%%%%%%%%%%%%%%%%%%%%%%%%%%%%%%%%%%%%%



\bibliographystyle{mnras}
\bibliography{paper2} % if your bibtex file is called example.bib

%%%%%%%%%%%%%%%%%%%%%%%%%%%%%%%%%%%%%%%%%%%%%%%%%%

%%%%%%%%%%%%%%%%% APPENDICES %%%%%%%%%%%%%%%%%%%%%

%\appendix



%%%%%%%%%%%%%%%%%%%%%%%%%%%%%%%%%%%%%%%%%%%%%%%%%%

% Don't change these lines
\bsp	% typesetting comment
\label{lastpage}
\end{document}